# Electric field-induced droplet deflection in microconfined flow

Somnath Santra[1], Shubhadeep Mandal[2], Aditya Bandopadhyay[1] and Suman Chakraborty[1,a]

[1]*Department of Mechanical Engineering, Indian Institute of Technology Kharagpur, Kharagpur, West Bengal - 721302, India*

[2]*Max Planck Institute for Dynamics and Self-Organization, Am Faßberg 17, D-37077 Göttingen, Germany*

The deflection of liquid droplet driven through a liquid medium under the combined action of transverse electric field and pressure driven flow has been studied in the present analysis. The present experimental and numerical analysis identifies the domain confinement as a key parameter for transverse migration of the droplets in the presence of a transverse electric field. Notably, the droplet migrates at a faster rate in highly confined domain. The present analysis also illustrates that the droplet can migrate toward the wall electrode or centerline depending on the physical and electrical properties of the system. The achieved steady state transverse position is found independent of its initial positions.

The dynamics of droplet in confined micro channel plays a significant role in interdisciplinary microfluidic and nanofluidic processes [1–3]. Its widespread applications include several chemical engineering [4–6] and bioengineering applications [6–9]. In these applications, a precise control over droplet motion is essential, allowing one to assess the influence of different actuating forces on the droplet motion in micro confinement. Typically, pressure driven flow has been used as an actuating mechanism for the generation and transportation of droplets. Typically, the migration of the droplet is influenced by the deformation of the interface [10–15], inertia of the fluid [16–18], fluid property [19–21], and surfactants [22–25], acoustic waves [26], magnetic fields [27] [28] and thermocapillary stresses [29,30]. Apart from these, droplet manipulation may be achieved by means of electricfields [31–38], where research communities exploit electric stress acting at the droplet surface for manipulation.

When a dielectric liquid droplet suspended in another dielectric liquid medium experiences electric field, electric stress is generated at the fluid-fluid interfaces due to the disparity of the electrical properties of the interacting fluids (Taylor-Melcher "leaky dielectric theory"). The electric stress leads the droplet to deform into prolate or oblate shape and creates a EHD flow circulation in and around the droplet [39,40]. The interface deformation and direction of EHD flow circulation depends on the electrical conductivity ratio $R=\sigma_i/\sigma_o$ and electric permittivity ratio $S=\varepsilon_i/\varepsilon_o$ of the droplets over the suspending fluid. When $R>S$, the droplet deforms into prolate shape and the flow circulations take place from the equators to the poles and an opposite scenario is observed for the cases when $S>R$. This EHD flow circulation creates a EHD force that causes the translation motion of the droplet when the droplet is placed in an off center position [41]. After the pioneering work of Taylor [39], several studies [42–51] have been made on the interaction of the droplet with the uniform or non-uniform electric field. Xu and Homsy [52] have studied the effect of uniform electric field applied along the axial direction on settling velocity and deformation of a leaky dielectric droplet suspended in another dielectric medium. In a recent study, Bandopadhyay et al. [53] have studied the lateral migration of a sedimentation droplet under a uniform tilted electric field. They have shown that the tilt angle has a significant impact on the direction and magnitude of the droplet velocity. Later on, Mandal et al. [54] have studied the cross stream migration of the droplet in unbounded plane poiseuille flow under a uniform electric field. They have shown that the droplet can migrate toward the centerline or wall electrode based on the tilt direction of electric field relative to the flow direction, and that for unbounded flows, mere transverse electric field is insufficient to cause transverse deviation from its trajectory.

[a] E-mail address for correspondence: suman@mech.iitkgp.ernet.in

Contrary to unbounded domains [54], we demonstrate experimentally that transverse electricfield has significant effect on the cross tream migration of the droplet in confined domain subjected to a back ground pressure driven flow. In spite of its significance relevance in several bio-fluidics and bio-medical process, this phenomenon has not been explored much till now.

The experimental set up as shown in Fig.1 that consists of a T shaped micro channel for the production of mono-dispersed droplet at desired frequency. The size of the droplet is smaller than the height of the channel (droplet radius, $a$~$0.30H$). The micro channel has one outlet port and three inlet ports. Primary inlet port is used for supplying the continuous fluid (c) whereas T-junction inlet is for the inflow of dispersed phase fluid (d). The properties of the interacting fluid are given in Table 1. After the production of the droplet at the T-junction, the droplet is migrated through the continuous fluid and at the diverging section, its position is deflected to an off center position by hydrodynamic lift produced by the flow from the secondary inlet port. At the diverging section, a uniform DC electric field is applied by a DC source meter (KEITHLEY-2410) along the transverse direction that modulates its migration characteristic. In the present analysis, we have considered two system: (a) system A, where droplets of silicon oil is in a medium of sun flower oil, (b) system B droplets of DI water is in a medium of silicon oil.

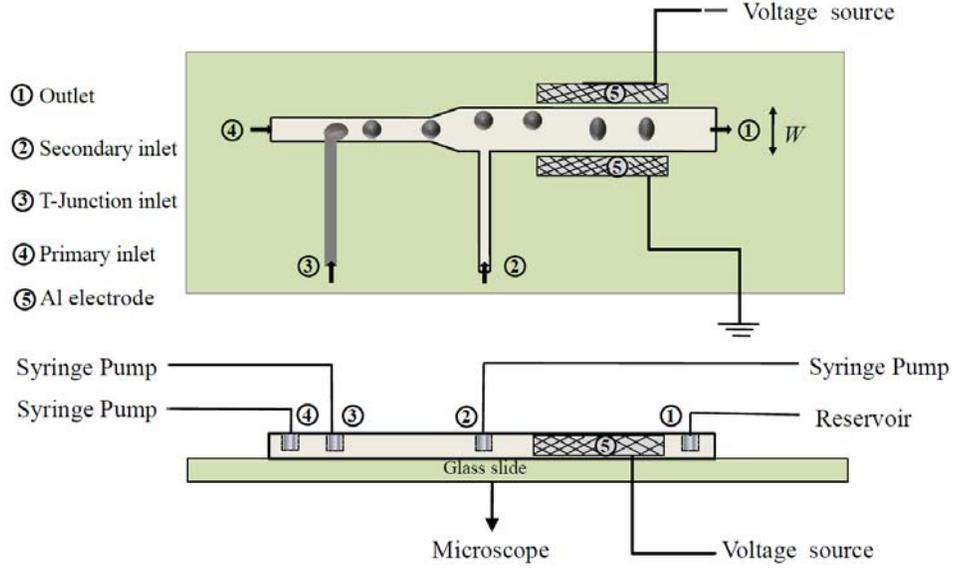

FIG. 1. A schematic of the experimental setup used for the present analysis.

For the fabrication of PDMS (poly-dimethyl siloxane) micro-channel, we have followed well established and standard methodology that consists of photo-lithography technique followed by soft lithography technique. Two thin aluminum electrode are placed on master pattern at the two sides of the channel wall at a specified distance from the wall before plasma bonding.

| Fluid | $\sigma$ (S/m) | $\varepsilon$ (F/m) | $\mu$ (Pa-s) | $\rho$ (Kg/m³) |
|---|---|---|---|---|
| DI water (d) | $5.49\times10^{-6}$ | $6.90\times10^{-10}$ | $1\times10^{-3}$ | 998 |
| Silicon oil (c/d) | $9.26\times10^{-11}$ | $3.43\times10^{-11}$ | $2.046\times10^{-2}$ | 1023 |
| Sunflower oil ( c ) | $4.74\times10^{-9}$ | $2.88\times10^{-11}$ | $4.9\times10^{-2}$ | 921 |

Table 1: Important properties of the fluids used.

Figure.2 shows the transient evolution of steady state transverse positions ($Y_d=y^*/W$, $y^*$ is the actual position of the droplet and $W$ is the width of the channel) of the droplet in confined microchannel. For system B having $S < R$, Fig.2(a) depicts that the droplet migrates faster towards the channel centerline with increases in the electric capillary number ($Ca_E=\varepsilon_e E_\infty^2 H/\gamma$, $E_\infty$ is the strength of the electricfield). Here $t$ is the time in non-dimensional form ($t = t^* u_c/H$, where $t^*$ is the actual time and $u_c$ is the characteristic velocity). On the other side, the figure also shows that the steady state position of the droplet shifts towards the wall electrode with increase in $Ca_E$ for system A having $S > R$. With the help of Fig.2(b), we have also shown the effect of domain confinement ($Wc=2a/W$) on droplet trajectory and it has been obtained that the migration of the droplet occurs at very faster rate with increase in the domain confinement. Along with the electric capillary number ($Ca_E$), one must acknowledge that the capillary number ($Ca=\mu_e u_c/\gamma$) as well as the electrical properties of the system also regulate the migration characteristics of the droplet.

In order to assess the complete physical picture of the EHD parameters, we resort to numerical results with the experimental results as shown in Fig. 2(a) and Fig. 2(b) and we have found good agreement between the experimental and numerical results. Furthermore, Fig. 3(a), Fig.3(b) and Fig.3(c) show excellent agreement between the numerically and experimentally obtained droplet shapes in confined domain. Motivated by the qualitative and quantitative results, we have further performed 2D numerical simulations believing that 2D motions will offer various realistic insights into the behavior of three-dimensional motion of the droplets and will uphold a considerable degree of physical relevance. Our consideration is also supported by the study of Stan et al. (2013) and Mortazavi and Tryggvason (2009). Through, numerical simulation, we have studied the effect of initial positions of the droplet on its steady state transverse position of the droplet as shown in Fig. (4) which shows that for the leaky dielectric system having $R>S$, the steady state position of the droplet is independent of its initial position and it always moves towards the centerline. But for a leaky dielectric system having $S>R$, we note that when the droplet is placed at the centerline, it moves straight towards the axial direction as depicted in Fig. 4(b). But when the droplet is placed at an off-center position above or below the centerline, it always achieves the same steady state transverse position above or below the centerline respectively and it does not depend on its initial positions. The insets of Fig.4(a) and 4(b) show the contour of deformed

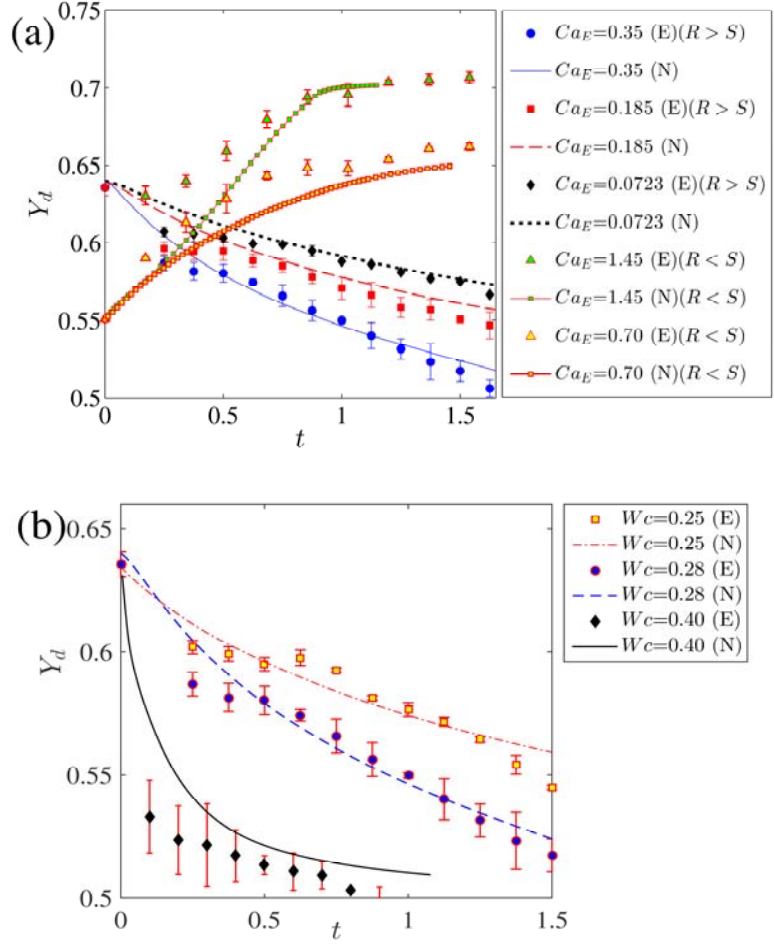

FIG. 2. Droplet trajectories for different values of (a) $Ca_E$ for system A and B (b) $Wc$ for system B with $Ca_E$ =0.35. For system A, others values are Ca=0.017, $(R, S)$=( 0.02, 1.22), $\lambda$ =0.42 and for system B, $(R, S)$=(O($10^4$), 20.1), $\lambda$ =0.05 Ca=0.0012, $Re \sim 10^{-2}$.

interface and direction of fluid flow circulation inside and outside of the droplet that determines the direction of EHD force (as discussed in the next section). Finally, we have constructed a region diagram shown in Fig. 5 clearly depicts three distinguishable regions based on the values of ($Ca$, $Ca_E$). The region containing 'pink triangular' data points indicates that the droplet migrates towards the channel wall, where as 'blue diamond' data points denote that the values of ($Ca$, $Ca_E$) for which droplet settles in between wall and channel centerline. The 'red triangular' data points represent that the droplet migrates towards the centerline

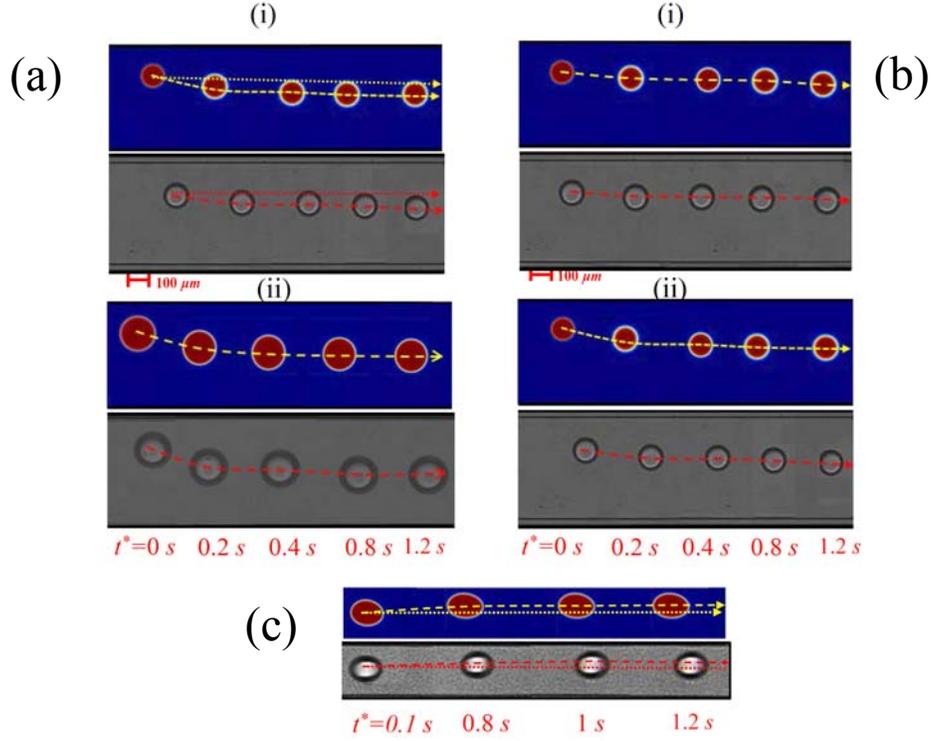

FIG.3. Images of the droplet at different time frame for (a) different electric field strength (i) $7\times10^2$ kV/m (ii) $9.58\times10^2$ kV/m. (b) different confinement ratio (i) $Wc=0.28$ (ii) $Wc=0.35$ for system B. Other parameters are $Y_d(t=0)=0.638$, $Ca=0.0012$, $Re\sim10^{-2}$. Fig. 3(c) shows the droplet shapes of system A for $E_\infty=13.07\times10^2$ kV/m, $Ca=0.017$, $Wc=0.52$ and $Y_d(t=0)=0.55$

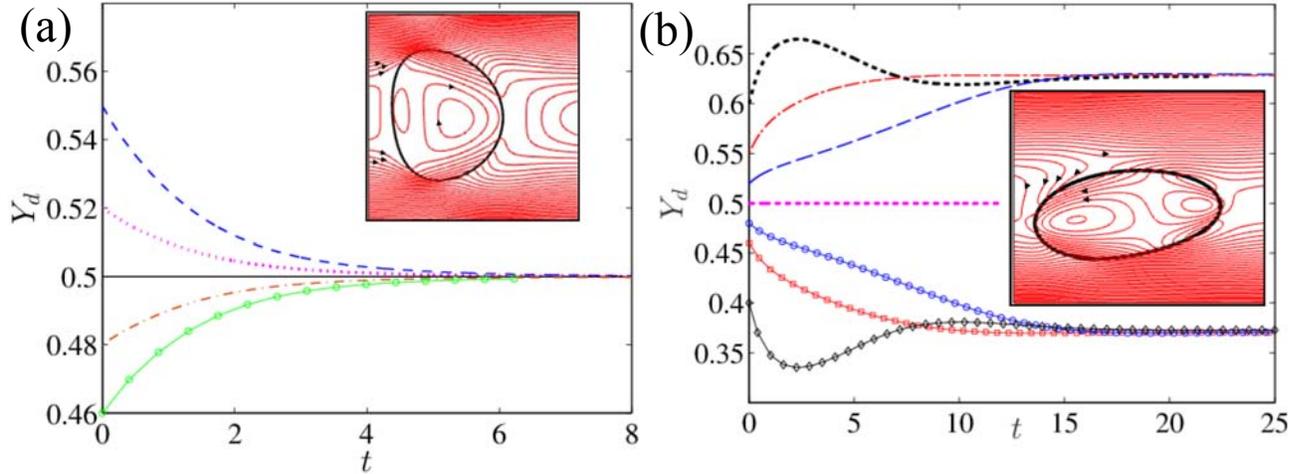

FIG 4. Effect of initial positions on the droplet trajectory for (a) system having $(S, R)=(0.5, 2)$ for $Ca_E=3.0$ (b) system having $(S, R)=(2, 0.5)$ for $Ca_E=2.0$. Others parameters are $Re=0.01$, $a=0.3$, $Ca=0.3$. The insets of the figures show the deformed droplet shape with fluid flow circulation

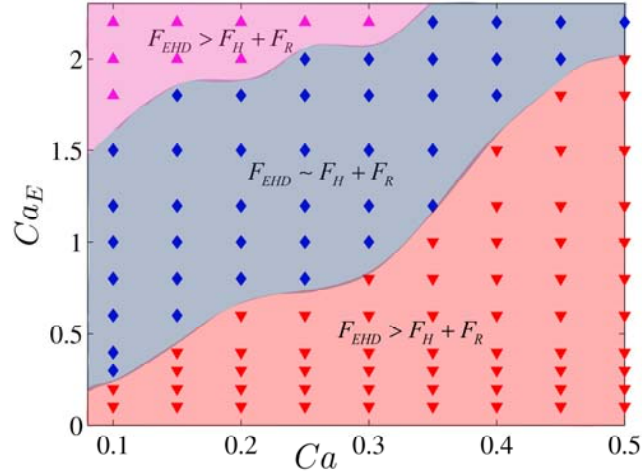

FIG.5. Regime diagram showing the three separate regions of droplet migration based on the values of ($Ca$, $Ca_E$). The other parameters are $Re$=0.01, $a$=0.3, ($S$, $R$)=(2, 0.5), $Y_d(t=0)$ =0.55.

For explaining the above observed phenomenon, we have taken four forces acting on droplet placed at an off center position. Those are: (i) hydrodynamic force ($F_H$) due to the non-uniform distribution of the velocity at the surface of the deformed droplet that leads the droplet to move towards the centerline, (ii) repulsive force ($F_R$) due to wall confinement called 'geometric blocking' [55] that also directs the motion of the droplet towards centerline and (iii) net EHD force ($F_{EHD}$) created due to the extra deformation caused by the electric field that can direct the droplet towards centre line or toward the wall electrode depending on the direction of flow circulation [41] and its strength is proportional to the strength of the electric field (or $Ca_E$). For system B having $R > S$, the direction of flow circulation takes place from equator to pole and $F_{EHD}$ directs the droplet towards the centerline. On contrary, for system A with $S > R$, the direction of flow takes place from pole to equator, and $F_{EHD}$ tries to move the droplet towards wall. At higher values of $Ca_E$, the higher magnitude of $F_{EHD}$ shifts the steady state position of the droplet towards wall electrode for system A, whereas it shifts droplet towards centerline at a very faster rate for system B. This phenomenon is reflected in Fig. 2(a). The reason behind the phenomenon observed in Fig 2(b) is also quite similar. Here, in addition to $F_H$ and $F_{EHD}$, $F_R$ is also prominent that causes faster migration. Furthermore, domain confinement induced enhanced electric field strength at the droplet surface also speeds up the cross stream migration via increasing the net $F_{EHD}$ force on the deformed droplet. For a leaky dielectric system having $S>R$, the $F_{EHD}$ tries to move the droplet towards the wall, where as $F_H$ and $F_R$ try to move towards the centerline. The relative strength of the involved forces determines the direction of the droplet movement and when $F_{EHD}$ is balanced by the combined strength of $F_R$ and $F_H$, the droplet attains a steady state transverse position in between the wall and centerline. This phenomenon is reflected in Fig.5. For a leaky dielectric droplet having $R>S$, the droplet always moved towards the centerline irrespective of its initial positions as all the involves forces are directed toward the centerline. But for $S>R$, $F_{EHD}$ is directed toward the wall electrode and in presence of high electric field strength, the droplet finds a same steady state transverse position in between wall electrode and centerline for different initial position.

In summary, through experimental and numerical investigation on the cross stream migration of a droplet in confined microchannel in combined presence of electricfield and background flow, we have identified that the domain confinement has considerable impact on the cross stream migration characteristic. Particularly, we have found that with increases in the domain confinement, the droplet migrates towards the centerline at a very faster rate. It is also interesting to note that based on the electrical properties, the droplets may migrate towards the centerline or towards the wall electrode and the migration occurs at faster rate with increase in electric field strength. Additionally, we have constructed a regime diagram that shows the dependency of migration characteristic on the *Ca$_E$* and *Ca*. Finally, our study denotes that for different initial positions, droplet surprisingly achieves same steady state transverse position. We hope that our experimental analysis not only reveals the physics underlying the droplet migration in confined domain but also portrays a novel way of modulating the droplet dynamics in confined domain that have several applications in modern days microfluidics devices.

## SUPPLEMENTARY MATERIAL

Supplementary material contains the details of numerical analysis and some videos of electric field modulated cross stream migration

## ACKNOWLEDGEMENT

SS is thankful to Mr. Tanoy Kahali for the insightful discussion on various experimental intricacies.